\documentstyle [11pt,epsfig]{article}
\def\ref{\bibitem{}}
\newcommand{\etal}{ {\it et al. }}
\newcommand{\ital}{ {\it et al., }}
\newcommand{\degg}{$^{\circ}$}
\newcommand{\cen}[1]{\centerline{#1}}
\begin{document}
\vspace*{3mm}
\cen{\large \bf Heliospheric Origin of Gamma-Ray Bursts} 
\footnote{submitted to {\sl Astrophysics and Space Science}} 
\begin{center}
Ti-Pei~~~Li  
\\{\sf High Energy Astrophysics Laboratory, Institute of High Energy Physics
\\Academy of Sciences, P.O.Box 918, Beijing, China}
\\Electronic mail: litp@{\scriptsize ASTROSV1.IHEP.AC.CN}
\end{center}
\vspace{4mm}
\parindent=0mm
\begin{quotation}
\cen{\bf Abstract} 
\vspace{3mm}
Systematic variations of average observational 
characteristics and correlation properties between different parameters 
of gamma-ray bursts (GRBs) with time from 1991 April to 1994 September 
are revealed. It is hard to explain the observed long-term variability 
by variations of experimental conditions. The variability of GRB properties
with the time scale of months to years, together with the similarity between
GRBs, solar hard X-ray flares and terrestrial gamma-ray flashes, may
indicate an origin of GRBs, at least partly, within the heliosphere.
Large-voltage and high-temperature pinch plasma columns produced by disruptive 
electrical discharges in the outer heliosphere 
can generate hard X-ray and $\gamma$-ray flashes with energy spectra
and spectral evolution characters consistent with that observed in GRBs.
\end{quotation}
\begin{quote}
\end{quote}
\vspace{4mm}
\parindent=5mm
\subsection*{1. Introduction}

  GRBs are intense, brief flashes of non-thermal hard X-rays and $\gamma$-rays
from random directions and at unpredictable times. GRBs have been extensively
studied since their discovery (Klebesadel, Strong \& Olson 1973). Several
satellites were launched to observe the bursts and a large number and wide
variety of models were put forwards to explain their origin.
The Burst and Transient Source Experiment (BATSE) on board the {\sl Compton 
Gamma Ray Observatory} (CGRO) was launched in April 1991. BATSE, with 
an unprecedented sensitivity, $\sim 1\times 10^{-7}$~erg cm$^{-2}$, has 
revolutionized the study about the nature of these objects and raised 
more debates on it as well. 

  After more than twenty five years of the discovery of GRBs, their origin still
remains a continuing puzzle. The most significant characteristic of these 
transient events is their complexity in morphology. The duration of GRB 
ranges from a few milliseconds to hundred of seconds and the temporal 
structure displays complicated patterns, of which basic features are still 
difficult 
to be summarized. In all timescales, some bursts have complex profiles, rich 
in fluctuating structures and some smooth, sometimes single structure profiles.
None of the models suggested so far for GRBs can provide
a satisfactory explanation to their main features in temporal and spectral 
characteristics.

 The observed bursts are distributed isotropically on the 
sky, that exclude distant galactic disk populations from the GRB sources.
The distributions of GRB intensities (peak counts $C$ or peak flux $P$),
integral $\lg N-\lg C$ or $\lg N-\lg P$ diagrams, are consistent with
the $-3/2$ slope (expected for a homogeneous sample of monoluminosity in
Euclidean space) for bright bursts, but considerably absent for bursts at 
low intensities. The incompatibility of the intensity distribution with
a uniform distribution throughout space rules out local galactic disk sources.
The only place left for GRBs at the Galaxy is at an extended galactic halo
with distances $d \approx 50-100$ kpc (Hartmannm 1994).   
The intensity distribution is widely accepted as a strong 
support to an extragalactic origin that the deviation from the $-3/2$ power law
is due to cosmological effects (Mao \& Paczy\'{n}ski 1992; Piran 1992; Dermer 
1992). The debate on the distance scale to GRBs is now mainly
between the galactic halo and cosmological distances (eg. Lamb 1995; 
Paczy\'{n}ski 1995). 

  Some average properties of GRBs detected by BATSE, e.g. duration and 
spectral hardness, show systematic variations with time from 1991 April to
1994 September (Li 1996). 
These variations are difficult to be explained with experimental effects. 
The analysis procedures and more results on the variability of 
properties of correlation between different GRB parameters are presented 
in section~2 of this paper in detail. The existence of GRB long-term 
variability with a time scale of months to years indicates a heliospheric 
origin. In section~3 a model of GRBs originating from disruptive discharges 
in instable plasma of the outer heliosphere is described. 
A study of two kinds of $\gamma$-ray flashes, solar flares and 
terrestrial $\gamma$-ray flashes, 
which were also detected by BATSE and certainly originated within the 
heliosphere, is made in comparison with GRBs in section~4.

\subsection*{2. Long-term variability}

  BATSE, having been continually detecting GRBs at a rate of about one 
event per day since April 1991, makes it possible to study long-term
variability of GRB observational properties.
The third BATSE (3B) Catalog of gamma-ray burst (Meegan \etal 1996) 
consists of 1122 $\gamma$-ray burst triggers and covered the time interval 
from 1991 April 21 until 1994 September 19.
To check through the 3B catalog to see if there exists systematic change with 
time in values of a burst parameter, we divided the observation time (TJD from 
8367 to 9614, where the truncated Julian Date TJD$=$JD$-$2440000.5) 
into a few epoch periods with the same length and calculated the average
value of the parameter for each period. 

\subsubsection*{2.1 Duration and hardness}

  The parameter $T_{90}$ is used to measure the duration of a burst.
$T_{90}$ is  the time interval in which the integrated counts from the burst 
increases from $5\%$ to $95\%$ of the total observed counts.
Firstly we calculated the average $\overline{T}$ and standard deviation 
$\sigma_{T}$ of all the values of $T_{90}$ listed in the 3B catalog, 
and then divided the observation time into four periods and 
for each of the four periods calculated the average, $<T_{90}>$, from all the 
$n$ bursts in each period for which $T_{90}$ are available. The result is shown 
in Fig.1(a). The statistical error of $<T_{90}>$ is estimated by 
$\sigma_{T}/\sqrt{n}$. One can see from Fig.1(a) that $<T_{90}>$ 
evidently decreases with time, from $41.4\pm3.6$ s during $91.4-92.2$ to only
$25.0\pm3.4$~s during $93.11-94.9$, at a rate , relative to 
$\overline{T}$, 
of $(-18.3\pm5.6)\%$~yr$^{-1}$, which was derived by a weighted linear 
least-squares fit with a reduced $\chi^{2}=0.21$ (the reduced $\chi^{2}$ is 
the fitting $\chi^{2}$ over the degrees of freedom). The linear regression
has $3.3\sigma$ significance for the null hypothesis that $T_{90}$
and observation time are uncorrelated. 

 The variability of $T_{90}$ has also been studied by the correlation 
analysis without binning. The correlation coefficient for a sample of 
$n$ pairs of $x,y$ values is defined as
\[r(x,y)=\sum_{i=1}^{n}\nu(i)/
\sqrt{\sum (x-\overline{x})^{2} \sum (y-\overline{y})^{2}}\]
 where 
$\nu(i)=(x(i)-\overline{x})(y(i)-\overline{y})$.
For the total $n=834$ bursts in 3B catalog for which
$T_{90}$ are available, the value of the correlation coefficient between
$T_{90}$ and the observation date $t$ was calculated
as $r(T_{90},t)=-0.12$. The significance of a value of $r$ to test the
hypothesis that $T_{90}$ and $t$ are uncorrelated can be evaluated by
the statistic \( u=r\sqrt{(n-2)/(1-r^{2})} \) which satisfies a
Student's $t$-distribution with $n-2$ degrees of freedom under the null
hypothesis (Siegel 1956).
We found $u=-3.4$. As $n=834\gg1$, the statistic $u$ here follows
the standard normal distribution, $u=-3.4$ indicates a result with
$3.4\sigma$ significance, which is consistent with the result from regression
analysis ($3.3\sigma$) and the probability that the observed variability
of $T_{90}$ comes from the statistical fluctuation is less than 
$3.2\times 10^{-4}$. 

\begin{figure}
\psfig{figure=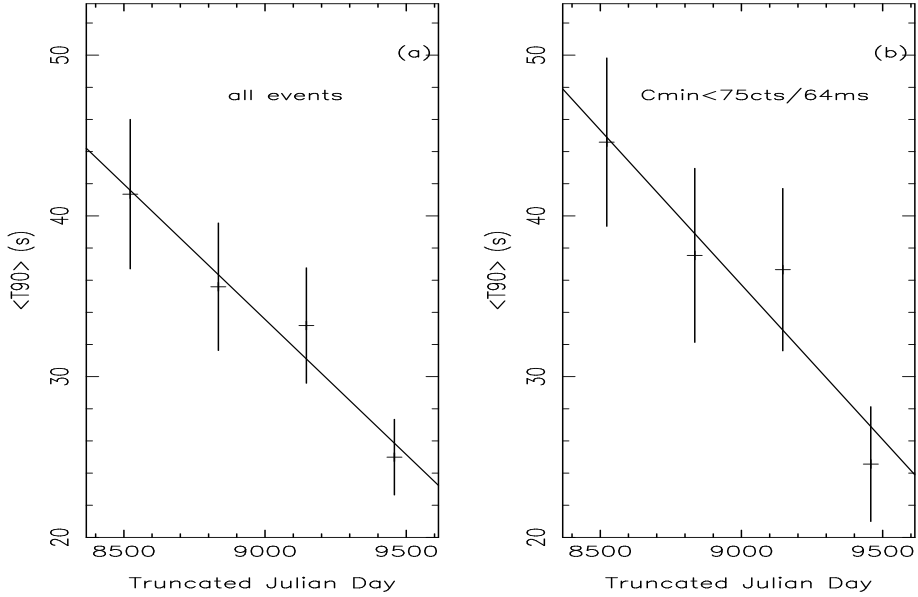,%
bbllx=180,bblly=18,bburx=600,bbury=594,%
width=7cm,height=16cm,angle=270}
\caption{Average burst duration $<T_{90}>$ against observation time
of 3B bursts.
(a) For all the bursts for which $T_{90}$ are available in 3B catalog; 
(b) For bursts in 3B catalog with trigger thresholds $C_{min}<75$ cts 
on the 64 ms timescale.} 
\end{figure}
  
  In the 3B catalog four energy fluences $F_{1}, F_{2}, F_{3}$ and $F_{4}$,
in units of erg cm$^{-2}$ and covering the energy ranges 20-50, 
50-100, 100-300 keV and $E>300$ keV respectively, are given for 867 bursts. 
The average fluences of bursts in each of four periods were calculated 
for the four
energy channels separately and shown in Fig.2(a). The fitted change slopes
from Fig.2(a) are listed in Table 1. One can see from Fig.2(a) and Table 1 
a steady trend that 
the average fluence increases quicker for higher energy channel.

\begin{figure}
\hbox{
\centerline{
\epsfig{figure=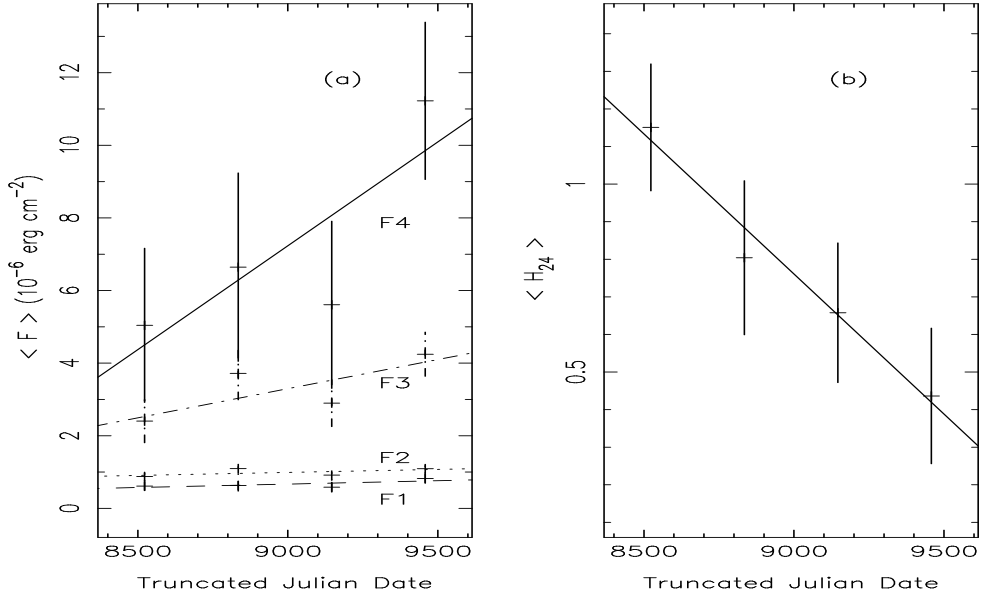,width=7cm,height=17cm,angle=270,%
bbllx=180,bblly=18,bburx=600,bbury=594}
}}
\caption{Evolution of burst spectral property. 
(a) Average energy fluences against observation time for 3B bursts. 
$F_{1}, F_{2},
F_{3}$ and $F_{4}$ are fluences covering the energy ranges 20-50,
50-100, 100-300 keV and $E>300$ keV respectively;
(b) Average spectral hardness ratio against observation time for 3B bursts. 
The hardness ratio $H_{24}=F_{2}/F_{4}$, 
from all the bursts for which $F_{2}$ and $F_{4}$ are available and 
$F_{2}/F_{4}<40$.}
\end{figure}

\begin{center}
Table 1: Change rates of average fluences\\
\vspace{4mm}
\begin{tabular}{c c c c c }   \cline{1-5} 
Energy range (keV) &  $20-50$ & $50-100$ & $100-300$ &$>300$ \\ \cline{1-5}  
Change rate $(\%$~yr$^{-1})$ & $10.2\pm10.2$ & $6.2\pm7.3$ & $17.8\pm9.7$ &
$29.0\pm$15.8\\  \cline{1-5}
\end{tabular}
\end{center}

  The ratio $H_{24}=F_{2}/F_{4}$ is used to measure spectral hardness. 
In statistics of $H_{24}$ a few events with very small $F_{4}$ 
seriously influence the average and disturb the general trend of variability 
much. Only three isolated bursts in total 690 bursts with $F_{4}>0$ have 
$H_{24}>40$, we picked them out and calculated $<H_{24}>$ and its standard 
deviation for residual 687 events for each period. 
The results shown in Fig.2(b) reveal that the energy spectra of GRBs
in 3B catalog became harder with time ($<H_{24}>$ decreased, from
$1.15\pm0.17$ during $91.4-92.2$ to only $0.44\pm0.18$ during $93.11-94.9$, 
at a rate of $(-35.1\pm11.9)\%$~yr$^{-1}$). 

\subsubsection*{2.2 Experimental effects}

  Since March of 1992 there are gaps in the data of GRB intensity, fluence or 
duration due to the errors of CGRO tape recorders, the problem has significantly
reduced since March of 1993 (Meegan \etal 1996). The number $n$ of events 
for which $T_{90}$ can be measured and that of total detected GRBs, $N$, 
for each period, are listed in Table 2. A less value of the ratio $n/N$ 
reflects greater effect of data gaps. As the change behaviour of $<T_{90}>$ 
and $<H_{24}>$, shown in Fig.1(a) and Fig.2(b), is quite different with that
of $n/N$, shown in Fig.3(a), it is hard to interpret the systematic 
variations of duration and hardness ratio by the effect of data gaps.   

The observed variation of burst duration can not be explained
by the variation of background either. The average of trigger thresholds
$C_{min}$ on the 64 ms timescale, set by command to a $5.5\sigma$ deviation
above background, was calculated for each period and also listed in Table 2 
and shown in Fig.3(b).
The values of $<C_{min}>$ have not shown a systematic change with time.

%\newpage
\begin{center}
Table 2: Statistics of 3B bursts in four periods\\
\vspace{4mm}
\begin{tabular}{c c c c c c c}   \cline{1-7} %\\
Period & \multicolumn{3}{c} {Number of Bursts} & $<C_{min}>$ & $<T_{90}>$ 
& $<H_{24}>$ \\
(yy.mm.dd) & $n$   & $N$ & $n/N$ & (cts/64ms) & (s) &    
\\ \cline{1-7} %\\
91.04.21--92.02.27 & 216 & 249 & 0.87 &69.9$\pm$1.5 &41.4$\pm$3.6 & 
1.15$\pm$0.17\\
92.02.28--93.01.04 & 168 & 271 & 0.62 &70.6$\pm$3.5 &35.6$\pm$4.1 & 
0.80$\pm$0.21\\
93.01.05--93.11.12 & 208 & 297 & 0.70 &66.7$\pm$0.9 &33.2$\pm$3.7 & 
0.66$\pm$0.19\\
93.11.13--94.09.19 & 242 & 305 & 0.79 &72.7$\pm$2.6 &25.0$\pm$3.4 & 
0.44$\pm$0.18\\ \cline{1-7}
\end{tabular}
\end{center}

\begin{figure} %Fig.3
\hbox{
\centerline{
\epsfig{figure=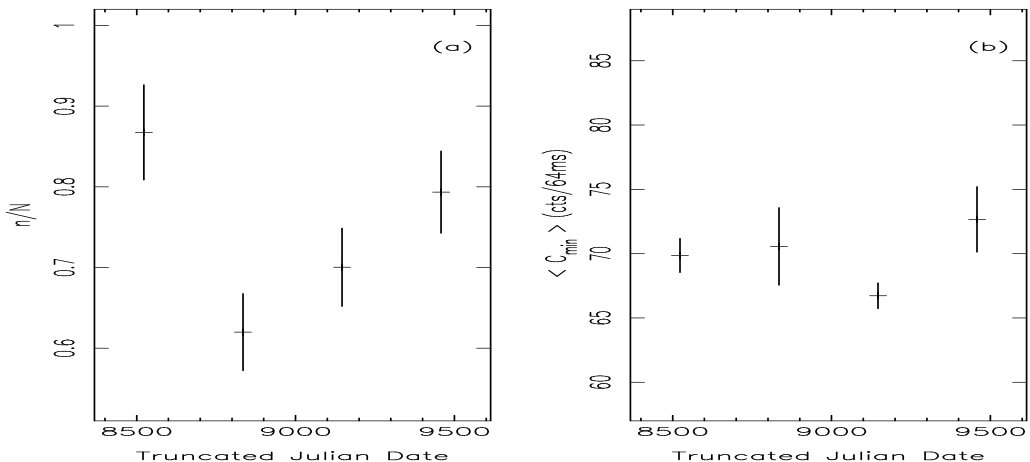,width=4.5cm,height=18cm,angle=270,%
bbllx=250,bblly=18,bburx=600,bbury=594}
}}
\caption{Variations of BATSE data gaps and backgrounds.
(a) Efficiency ratio $n/N$ for measurement of $T_{90}$ against observation 
time, where $N$ is the total number of detected bursts during a period
and $n$ the number of bursts for which $T_{90}$ can be measured during 
the same period.
(b) Average trigger threshold on the 64 ms timescale against observation time.}
\end{figure}

  For further inspecting the effect of background variation on $<T_{90}>$, 
the $C_{min}$ distribution for bursts in the 3B catalog 
was made and shown in Fig.4(a). Most values of $C_{min}$ are smaller than
75 cts/64ms. The bursts were grouped into four subsets with $C_{min}<65, 
65\leq C_{min}<68, 68\leq C_{min}<75$ and $C_{min}\geq 75$ (in units of 
cts/64ms) respectively. 
The average values of $T_{90}$ for each subsets were calculated and shown 
in Fig.4(b).There is no visible dependence between $<T_{90}>$ and $C_{min}$ 
except such detected bursts under highest trigger thresholds of 
$C_{min}\geq75$ cts/64ms. Fig.1(b) shows that the average duration $<T_{90}>$ 
for bursts having $C_{min}<75$ cts/64ms also decreases with time at a rate of
$(-17.1\pm5.2)\%$~yr$^{-1}$, consistent with that derived from all bursts.
Therefore, the observed duration variability can not be interpreted with 
background variation. 

\begin{figure} %Fig.4
\hbox{
\centerline{
\epsfig{figure=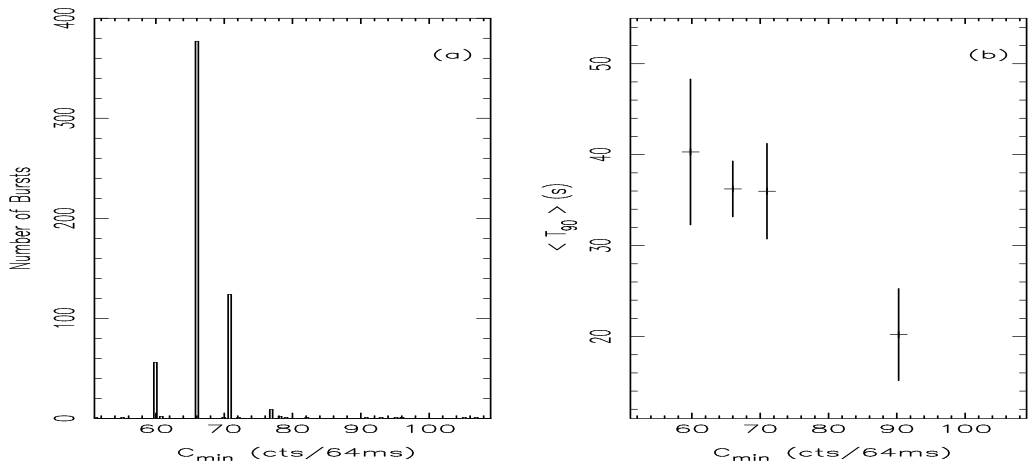,width=6cm,height=18cm,angle=270,%
bbllx=120,bblly=18,bburx=600,bbury=594}
}}
\caption{Distribution of trigger thresholds on the 64 ms time scale
of 3B catalog;
(b) Correlation between duration and trigger threshold of 3B bursts.}

\end{figure}

  Another experimental condition with secular change is spacecraft altitude. 
Fig.5 shows CGRO altitude history. After comparing Fig.5 with Fig.1 and Fig.2, 
one can see that the observed long-term variations of duration and hardness
can not be due to the altitude variation either.

\begin{figure} %Fig.5
\hbox{
\centerline{
\epsfig{figure=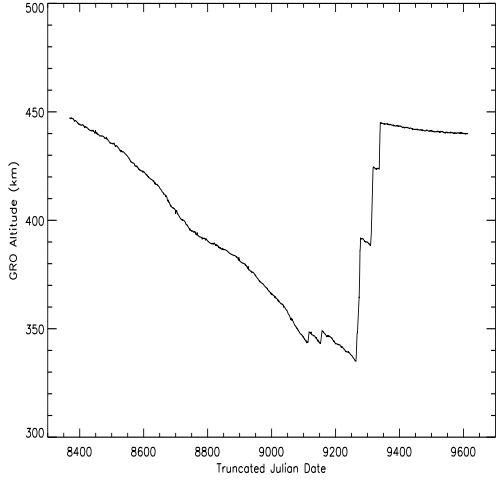,width=6cm,height=7cm,angle=90, %
bbllx=54,bblly=54,bburx=500,bbury=738}
%}}
\parbox[b]{8cm}{
\caption{CGRO altitude vs. time.}}
}}
\end{figure}

  There exists a positive dependence of burst fluences against their durations.
Fig.6(a) shows the correlation of $\lg F$ and $\lg <T_{90}>$ in 3B events
where the regression line has a slope of $\lambda = 0.52\pm0.02 $. 
If the decrease of average duration is caused by decrease of detection 
efficiency for longer events by some unknown reasons, a decrease of average 
fluence with time should 
be expected. But the measured results for fluences show an opposite trend, see 
Fig.6(b), indicating that the detected changes should be intrinsic.

\begin{figure} %Fig.6
\hbox{
\centerline{
\epsfig{figure=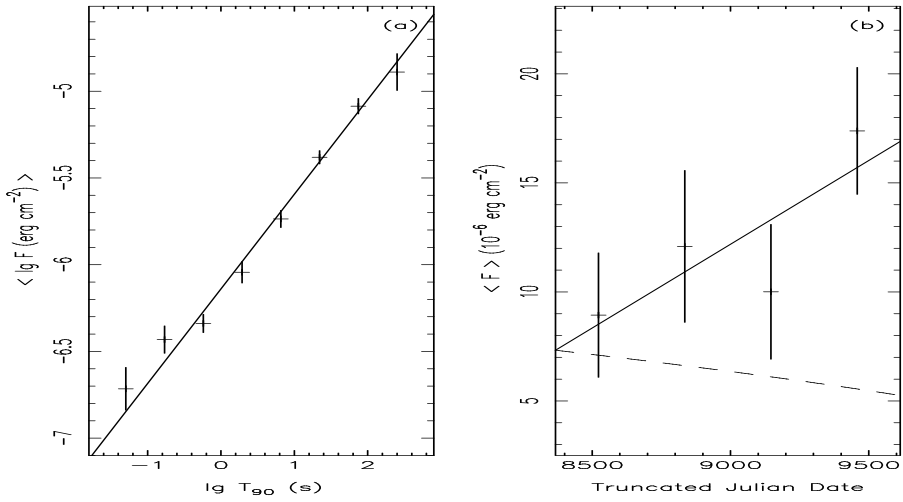,width=7cm,height=18cm,angle=270,%
bbllx=25,bblly=18,bburx=550,bbury=594}
}}
\caption{(a) Correlation between fluence and duration of 3B bursts;
(b) Average fluence against observation time, solid lines represent the
measured results and dashed line is the expectation calculated from the 
regression line for $<T_{90}>$ in Fig.1(a) and correlation between 
$F$ and $T_{90}$ and normalized at TJD=8367.}  
\end{figure}

\subsubsection*{2.3 Correlation properties}

  The relationship between two parameters of GRBs reflects the nature
of GRB production process. Long-term variability also found for
different correlation property of GRBs in 3B catalog, which is more
difficult to be interpreted with experimental effects (Li 1997). 
There are total 807 bursts in the 3B catalog for which both $F$ and $T_{90}$ 
are available. Calculating the correlation coefficient between $\lg F$ and 
$\lg T_{90}$ for the 807 bursts we get $r(\lg F,\lg T_{90})=0.67\pm0.02$. 
The statistical 
error of the correlation coefficient is estimated by the bootstrap 
technique (Efron 1979; Diaconis \& Efron 1983). To check through the 3B 
catalog to see if there exists systematic change with time in the relationship 
between $F$ and $T_{90}$, we divide the observation time $t$ (TJD from 8367 
to 9614 for 3B catalog) into five epoch periods, letting each period has 
nearly equal number of bursts for which both $F$ and $T_{90}$ are available, 
and calculate the correlation coefficient $r(\lg F, \lg T_{90})$ and 
its statistical error for each period. Fig.7(a) shows the result.
From Fig.7(a) one can see that the correlation coefficient 
$r(\lg F, \lg T_{90})$ increased from $0.60\pm0.04$
during $91.111 - 91.329$ to $0.71\pm0.03$ during $94.154 - 94.262$ at a rate
$\lambda(r,t)=0.038\pm0.017$ yr$^{-1}$, which was derived by a weighted linear
least-squares fit. The linear increasing
of correlation coefficient $r(\lg F, \lg T_{90})$ with observation time $t$ has 
$2.2\sigma$ significance for the null hypothesis that $r(\lg F, \lg T_{90})$ 
and observation time $t$ are uncorrelated.

 Now we study the variability of correlation coefficient between spectral 
hardness and duration. The ratio $H_{32}=F_{3}/F_{2}$ is used to measure 
spectral hardness. 
As events with very small $F_{2}$ considerably disturb the general trend 
of variability of $H_{32}$, we pick a couple of events with $H_{32}>10$ out in
the statistics through this work. There is a general trend of hardness 
with short events 
being harder (Dezalay \etal 1992; Kouveliotou \etal 1993a), but the
variation behaviour of hardness against duration for short bursts is quite 
different with that for long bursts:  the correlation coefficient 
$r(H_{32}, \lg T_{90})$ 
between $H_{32}$ and $\lg T_{90}$  for 583 long bursts with $T_{90}>2$ s
in the 3B catalog is 0.01 and that for 178 short bursts with $T_{90}<2$ s is 
$-0.11$.
No visible change of the correlation coefficient between $H_{32}$ and 
$\lg T_{90}$
with time has been found for long bursts, the slope of the regression line 
of $r(H_{32}, \lg T_{90})$ on $t$ fitted to the values of five epochs 
being $0.005\pm0.039$ yr$^{-1}$, whereas     
evident variation existed in short bursts, shown in Fig.7(b).
The correlation coefficient $r(H_{32},\lg T_{90})$ for bursts with 
$T_{90}<2$ s increased
from $-0.52\pm0.11$ during $91.111 - 91.324$  to $0.16\pm0.13$ during
$94.43 - 94.262$. The regression line in Fig.7(b) has a slope of
$0.239\pm0.057$ yr$^{-1}$, showing a $4.2\sigma$ long-term variability.
        
\begin{figure}
\hbox{
\centerline{
\epsfig{figure=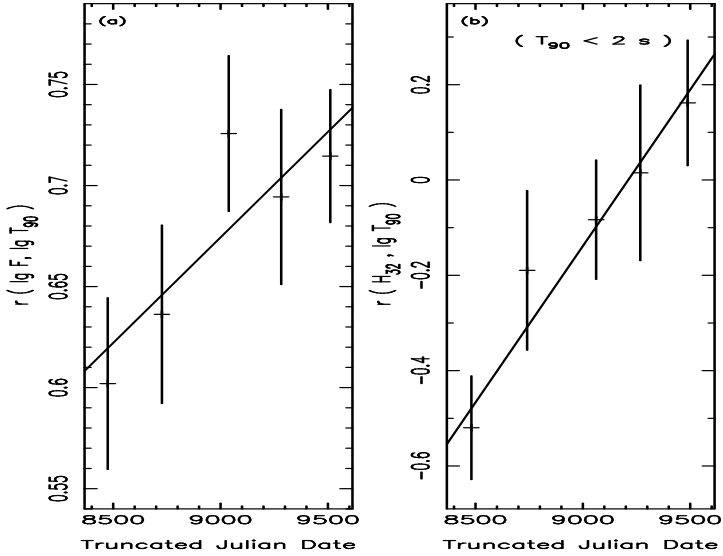,width=6cm,height=18cm,angle=270,%
bbllx=200,bblly=18,bburx=600,bbury=590}
}}
\caption{Evolution of correlation coefficients. 
(a) Correlation coefficient between $\lg F$ and $\lg T_{90}$ 
against observation time for 3B bursts. 
(b) Correlation coefficient between hardness ration $H_{32}$ and 
$\lg T_{90}$ against observation time for short 3B bursts with 
$T_{90}<2$ s.} 
\end{figure}

\subsubsection*{2.4 Bimodality}

  The bursts duration distribution can be divided to two sub-groups according 
to $T_{90}$: Long bursts with $T_{90}>2$ s and short bursts with $T_{90}<2$ s
(Kouveliotou \etal 1993b; Meegan \etal 1996). We found that the variability 
behaviour of observed GRB parameters is also bimodal: duration becoming 
shorter and spectrum harder is only 
observed for long bursts in the 3B catalog, but not for short ones. 
On the other hand, the variability of correlation properties between different
parameters, e.g. $F$ and $T_{90}$, or $H_{32}$ and $T_{90}$,
is revealed in short bursts more obviously than that in long bursts.
Fig.8 show time variation behaviour of the parameters (duration, 
hardness ratio and fluence), and the correlation coefficients  
between fluence and duration and between hardness and duration
for long bursts and for short bursts in the 3B bursts separately. 
The change slopes derived by the linear fitting procedure are listed in Table 4.
For short bursts both $r(\lg F, \lg T_{90})$ and $r(H_{32}, \lg T_{90})$
increased with time evidently with $4\sigma$ significance, but for long bursts
the change rate of $r(\lg F, \lg T_{90})$ was only one third of that for short
bursts and no visible change found for $r(H_{32}, \lg T_{90})$. 

\begin{figure} %Fig.8
\hbox{
\centerline{
\epsfig{figure=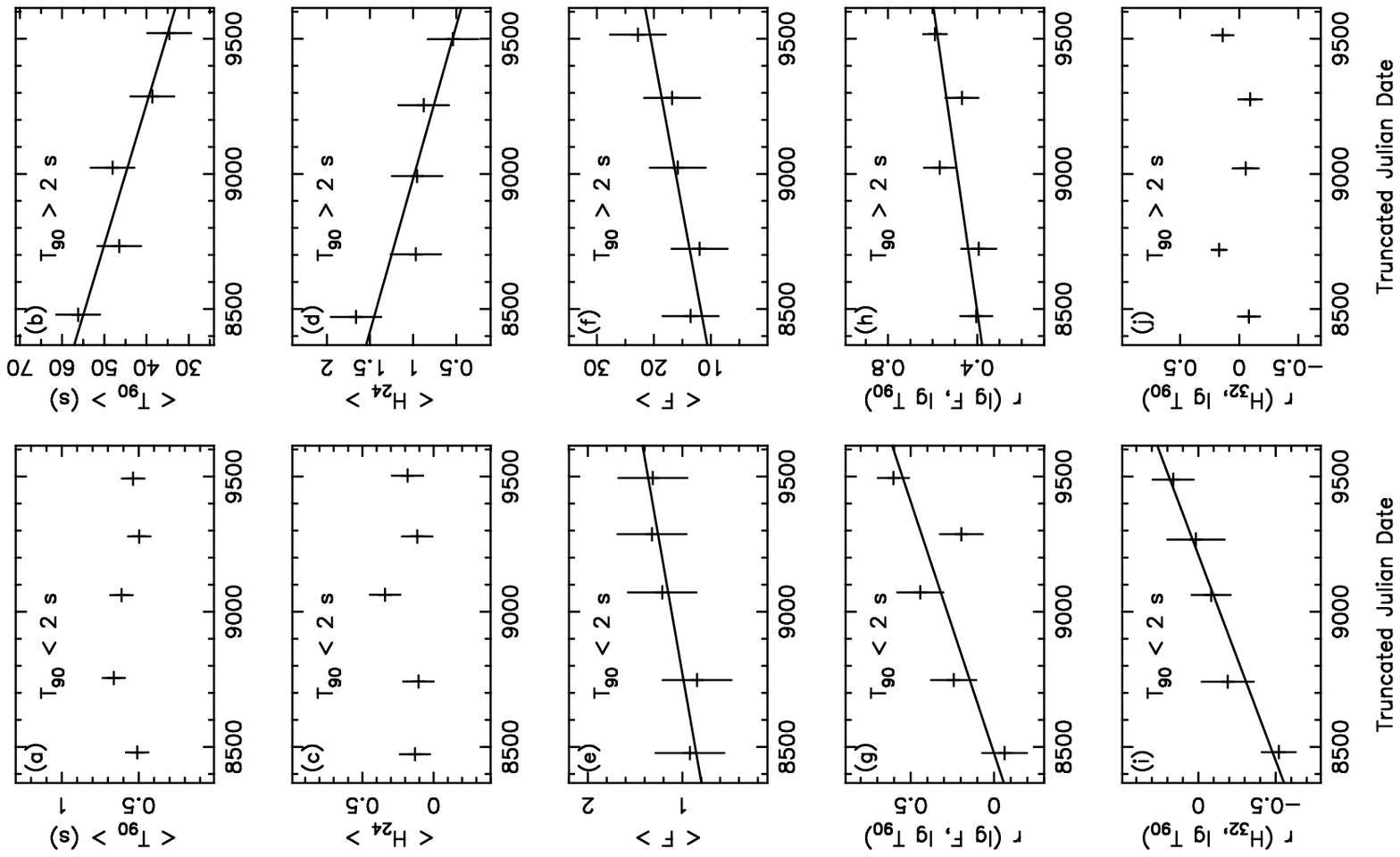,width=20cm,height=19cm,angle=270,%
bbllx=60,bblly=18,bburx=550,bbury=404}
}}
\caption{Evolution of average values of parameters $T_{90}, H_{24},
F$ and correlation coefficients $r(\lg F, \lg T_{90}), r(H_{32}, \lg T_{90})$ 
for two types of GRBs in the 3B catalog.
(a),(c),(e),(g),(i) for bursts with $T_{90}<2$ s, (b),(d),(f),(h),(j) 
for bursts with $T_{90}>2$ s. 
$H_{24}=F_{2}/F_{4}$ and $H_{32}=F_{3}/F_{2}$. A couple of bursts with 
$H_{24}>40$ 
were excluded from the statistics in (c) and (d), and that with $H_{32}>10$ 
in (i) and (j). } 
\end{figure}
\newpage
\begin{center}
Table 3: Change slopes of GRB properties \\
\vspace{4mm}
\begin{tabular}{c c c}   \cline{1-3} 
Property & \multicolumn{2}{c} {Change slope (yr$^{-1})$} \\
 & short bursts ($T_{90}<2$ s) & long bursts ($T_{90}>2$ s) 
\\ \cline{1-3} %\\
$T_{90}$          & $(-2.5\pm5.8)\%$  & $(-15.7\pm5.0)\%$ \\
$H_{24}$          & $(12.0\pm27.2)\%$ & $(-32.3\pm12.9)\%$ \\
$F$               & $(16.3\pm14.6)\%$ & $(19.7\pm13.2)\%$ \\
$r(\lg F, \lg T_{90})$   & $0.20\pm0.05$     & $0.064\pm0.028$     \\
$r(H_{32}, \lg T_{90})$   & $0.24\pm0.06$     & $0.005\pm0.039$     \\
\cline{1-3}
\end{tabular}
\end{center}

\subsubsection*{2.5 Short bursts}

  From Fig.8(i) and (j) and Table 3 one can see that long-term variability
of the correlation coefficient $r(H_{32}, \lg T_{90})$ only show in short bursts
with $T_{90}<2$ s, none in long bursts. Now we further inspect the correlation
properties of short bursts.  

  There are 745 events in the 3B catalog for which both $T_{90}$ and $F$ are 
available. To show the relationship between hardness and duration, we divide
the range $0 - 200$ s of $T_{90}$ into 22 bins, letting each bin includes 
nearly equal number of events, and calculate the average of $T_{90}$,
average of $H_{32}$ and their standard deviations for each bin. Fig.9 shows
the result. It is noticeable in Fig.9 that a turn in the correlation behaviour 
between $H_{32}$ and $T_{90}$ occurs at $T_{90}\simeq 0.55$ s as well as 
at 2 s.  
%\newpage
\begin{figure}   %Fig.9
\hbox{
\centerline{
\epsfig{figure=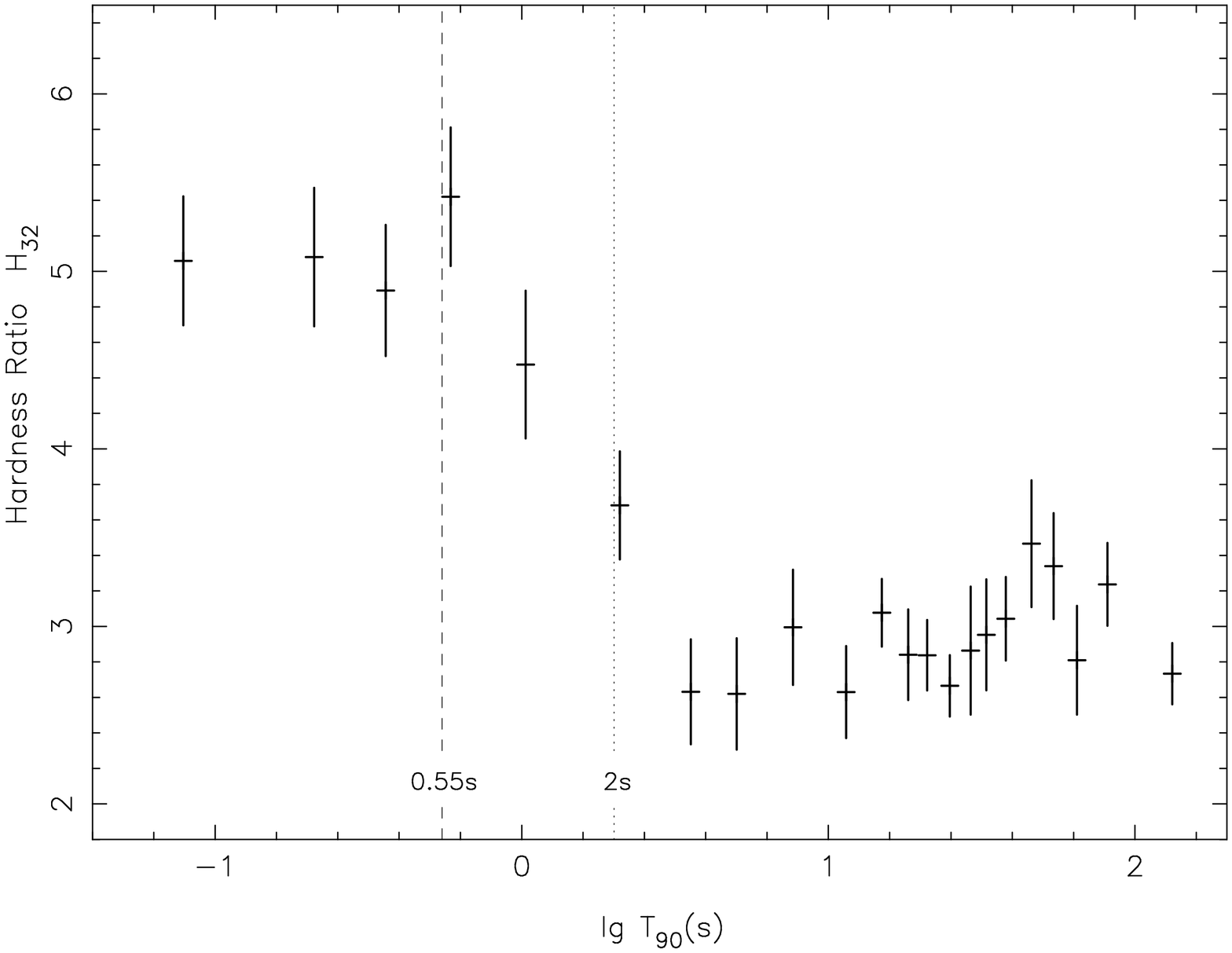,width=6.2cm,height=6.35cm,angle=270,%
bbllx=18,bblly=230,bburx=600,bbury=594}
}}
\caption{Distribution of $H_{32} - \lg T_{90}$ for all the events in 3B 
catalog during the period between 8367 and 9614 TJD. }   
%\end{figure}
%\begin{figure}    %Fig.10
\hbox{
\centerline{
\epsfig{figure=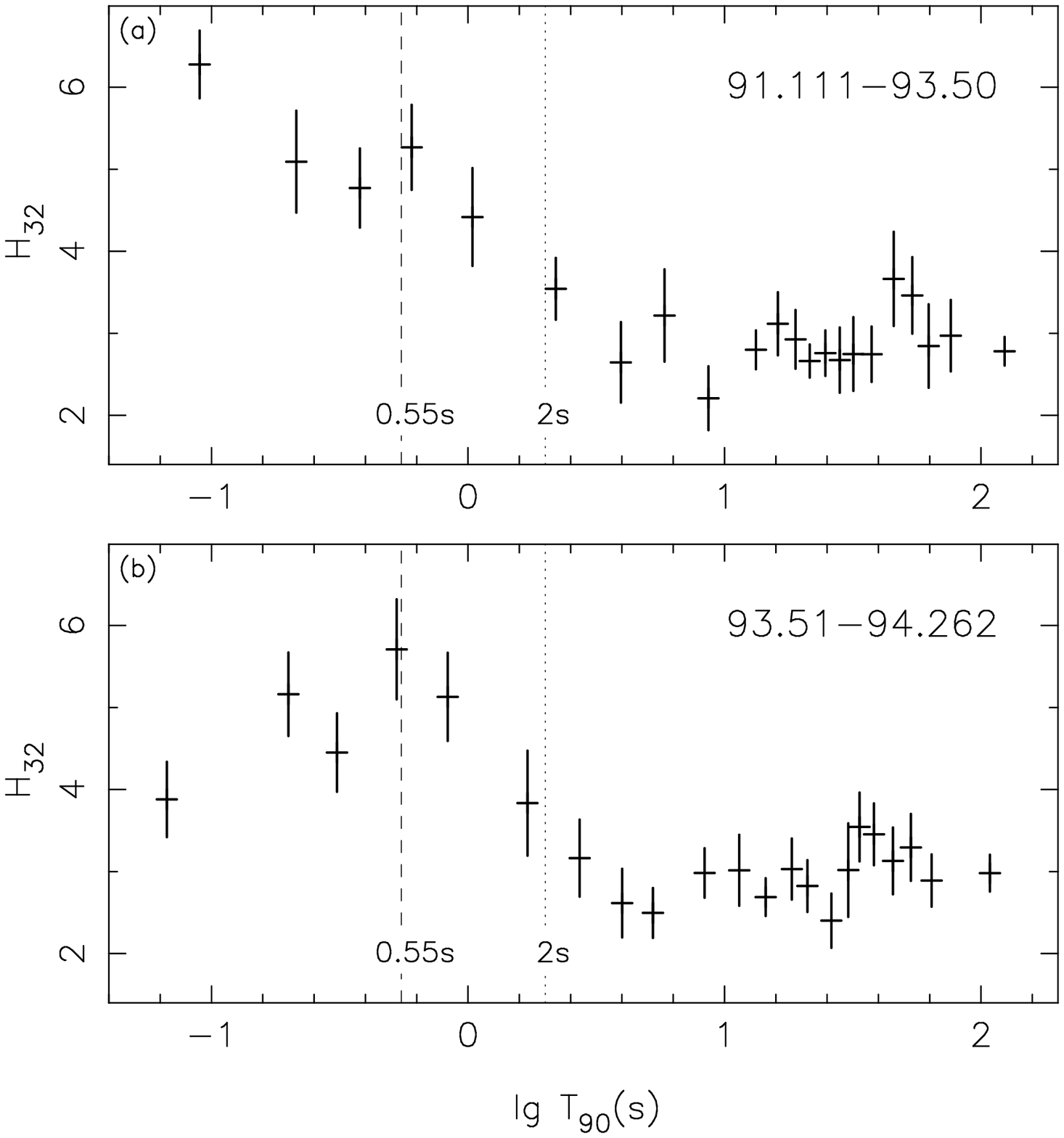,width=12cm,height=13.5cm,angle=270,%
bbllx=18,bblly=18,bburx=600,bbury=594}
}}
\caption{$H_{32} - \lg T_{90}$ distributions. (a) For bursts during
the period between 8367 and 9037 TJD. (b) For bursts during the period
between 9038 and 9614 TJD. }
\end{figure}
From Figure 10 (a) and (b), which show the $H_{32} - \lg T_{90}$ distribution
of 373 bursts during observation epoch $t$ between 8367 and 9037 TJD
and that between 9038 and 9614 TJD separately, one can see that the correlation
behaviour between $H_{32}$ and $T_{90}$ of the brief bursts 
($T_{90}<0.55$s) is evidently changed with time. 
 
  Similar phenomena can also be found in the correlations between $H_{41}$ 
and $T_{90}$ and that between the peak flux $P_{64}$, 
the maximum flux in $50-300$ keV 
(ph. cm$^{-2}$s$^{-1}$) integrated over 64 ms timescale, and $T_{90}$ 
for the short bursts.
The correlation coefficient $r(H_{41},\lg T_{90})$ evidently increased with 
observation time at a rate $\lambda=0.36\pm0.08$ yr$^{-1}$ for the brief 
bursts shown in Fig.11(a), but  no visible variation
is found in Fig.11(b) for the medium bursts ($\lambda=0.008\pm0.12$ yr$^{-1}$).
For the brief bursts the correlation coefficient of $r(P_{64},\lg T_{90})$ is 
positively correlated with the observation time (the slope of regression line
$\lambda=0.26\pm0.06$ yr$^{-1}$ in Fig.11(c)), but for the medium bursts
a negative correlation between $r(P_{64},\lg T_{90})$ and the
observation time appears with $\lambda=-0.15\pm0.10$ yr$^{-1}$ 
shown in Fig.11(d).   

\begin{figure}  %Fig.11
\hbox{
\centerline{
\epsfig{figure=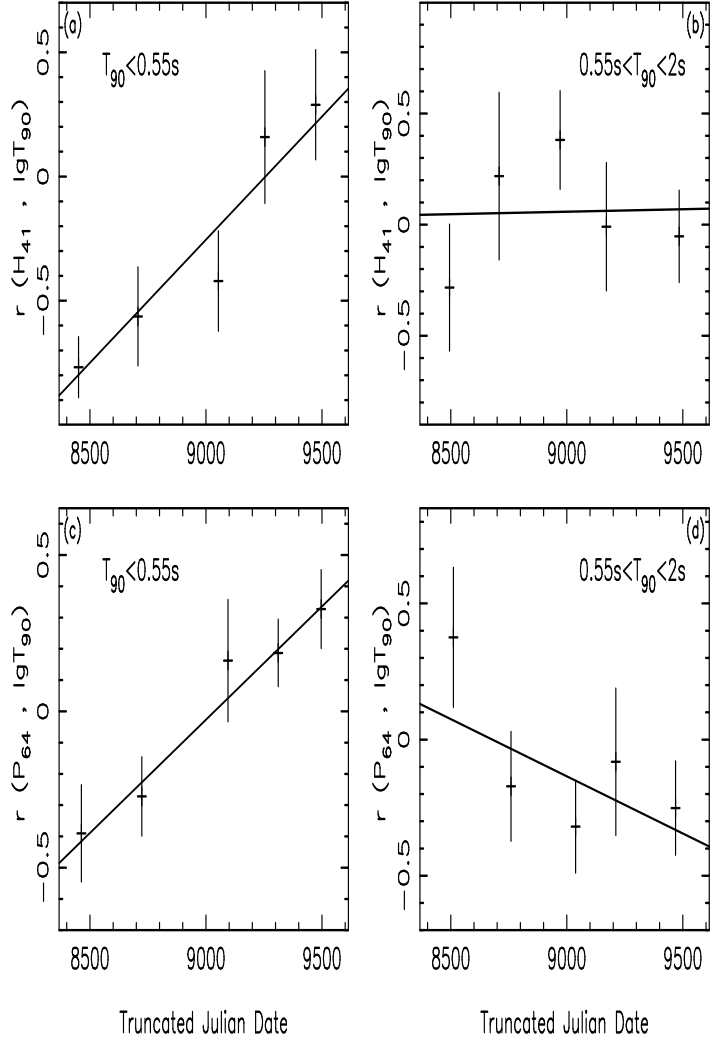,width=17cm,height=5cm,angle=270,%
bbllx=18,bblly=250,bburx=600,bbury=594}
}}
\caption{Correlation coefficient vs. observation time. For each figure the 
observation time period of 3B catalog is divided into five epoch periods 
by letting each period has nearly equal number of available events.
The correlation coefficient $r$ is calculated for each bin and its standard
deviation is estimated by the bootstrap technique. A couple of isolated events
with relevant parameter $H_{41}>150$ or $P_{64}>15$ are excluded from 
statistics. 
(a) $r(H_{41},\lg T_{90})$ vs. observation epoch for $T_{90}<0.55$s.  
(b) $r(H_{41},\lg T_{90})$ vs. observation epoch for 0.55s$<T_{90}<2$s.  
(c) $r(P_{64},\lg T_{90})$ vs. observation epoch for $T_{90}<0.55$s.  
(d) $r(P_{64},\lg T_{90})$ vs. observation epoch for 0.55s$<T_{90}<2$s.} 
\end{figure}

   To get a free of binning estimation of the significance level 
for the time variability of the correlation between $x$ and $y$
from a sample of $n$ points $(x,y,t)$, the equation of the line 
of regression of $x$ on $t$, $x=A_{x}+B_{x}t$, and that of $y$ on $t$, 
$y=A_{y}+B_{y}t$, are firstly derived by the method of least squares.
For each burst $i$ $(i=1,...,n)$ with parameters $(x(i), y(i), t(i))$
the following quantity $\nu(i)=(x(i)-x_{0}(i))(y(i)-y_{0}(i))$ with
$x_{0}(i)=A_{x}+B_{x}t(i)$ and $y_{0}(i)=A_{y}+B_{y}t(i)$ is calculated.
The correlation coefficient between $\nu$ and $t$, $r(\nu,t)$, for the sample 
of $n$ pairs of $\nu, t$-values now can be calculated, the standard deviation
of the derived $r(\nu,t)$ can be evaluated by the bootstrap technique
and the relative portion $\xi$ of the correlation
coefficient less or equal to 0 (or greater or equal to 0  if
$r(\nu,t)<0$) among the bootstrap samples can be taken as an estimate 
of the significance level.
The obtained values of $r(\nu,t)$ of the parameter pairs
$(H_{32}, \lg T_{90}), (H_{41}, \lg T_{90})$, and $(P_{64}, \lg T_{90})$ 
are listed in Table 4 for the two types of burst separately, where the standard
deviations and significance levels are derived from $10^{7}$ 
bootstrap samples for each case. 

%\newpage 
\begin{center}
Table 4: Significance levels $\xi$ of correlation variations\\

\vspace{4mm}
\begin{tabular}{c c c c c }   \cline{1-5} %\\
Correlated Parameters & \multicolumn{2}{c} {$T_{90}<0.55$s} & 
\multicolumn{2}{c} {0.55s$<T_{90}<2$s} \\
(x, y) & $r(\nu,t)$   & $\xi$ & $r(\nu,t)$ & $\xi$     
\\ \cline{1-5} %\\
$H_{32},~ \lg T_{90}$ & $0.30\pm0.09$ & $2.1\times10^{-4}$ & $-0.03\pm0.10$
 & 0.38 \\
$H_{41},~ \lg T_{90}$ & $0.39\pm0.07$ & $9.3\times10^{-5}$ & $-0.07\pm0.11$ &
0.26 \\
$P_{64},~ \lg T_{90}$ & $0.36\pm0.07$ & $4.8\times10^{-6}$ & $-0.16\pm0.09$ &
$4.1\times10^{-2}$\\ \cline{1-5}
\end{tabular}
\end{center}

   One can see that the three correlation coefficients listed in Table 4
have quite different variation behaviour between the two types of burst and
significant (near to or great than $4\sigma$)
variations appear in the correlation properties of the brief bursts
($T_{90}<0.55$ s).

\subsection*{3. A discharge model}

Alfv\'{e}n (1981) has stressed the importance of studying electric current
circuits in plasmas to understanding phenomena occurred in the
magnetosphere till galactic dimensions, and held that many of the explosive
events observed in cosmic physics are produced by exploding electric double
layers. In a circuit with inductance $L$ and current $I$ in plasma, 
electric double
layers can be produced, in which large potential drops may be built up
over distances of the order of some tens of the Debye lengths. A disruption
of the current in a layer by plasma instability will cause an explosion:
the layer voltage, $V=L(dI/dt)$, may exceed the normal value by
several orders of magnitude and the magnetic energy stored in the circuit 
being suddenly released in the layer, will cause the plasma
column to collapse to a much smaller radius by inward magnetic pressure of
the discharging current (Z pinch) (Krall \& Trivelpiece 1973) and the 
resulting compressed,
large-voltage and high-temperature discharge column will radiate energetic
photons from the inverse Compton scattering between the high-energy electrons 
and thermal photons. In an isotropic thermal emission with mean photon
energy $\overline{h\nu}$, the mean energy of a
scattered photon from an electron with an energy $\epsilon$ is 
$\overline{E}=\frac{4}{3}(\frac{\epsilon}{mc^{2}})^{2}\overline{h\nu}$
(Ginzburg \& Syrovatskii 1964).
Let the temperature of a pinch discharge column is $T$ (K), the differential 
photon density of the thermal emission at frequency $\nu$ is
$n_{\nu}(T)=B_{\nu}(T)[1-\exp(-\tau(\nu))] $,
where $B_{\nu}(T)$ is the photon density of a black body and
the optical depth 
\[ \tau(\nu)=4.1\times 10^{-23}T^{-3.5}(\frac{h\nu}{kT})^{-3}
(1-e^{-\frac{h\nu}{kT}})\bar{g}\int N^{2}dl \]
with the emission measure $\int N^{2}dl$ being in cm$^{-5}$ and average Gaunt
factor $\bar{g}$ taken as the approximation (Kulsrud 1954)
\[ \bar{g}=\frac{\sqrt{3}}{\pi}\ln\frac{\sqrt{0.76+h\nu/kT}+0.87}{\sqrt{0.76+
h\nu/kT}-0.87} \]
For thermal emissions of temperature $T=10^{6}$ K and optical depth 
$\tau_{0}=0.1$ at $h\nu=1kT$, the mean photon energy $\sim 72$ eV,
then the mean energy of a scattered photon from an electron with $\epsilon=100$
MeV can be calculated as $\sim 3.7\times 10^{3}$ keV. Therefore, the large 
voltage and high temperature discharge column is an effective radiator 
of hard X-rays and soft $\gamma$-rays.

  The existence of any intrinsic variability of GRB properties with a time scale
of months to years can not be interpreted with either the galactic or cosmological 
origin. Only the decrease in solar activity during the 3B observation period 
remains to be a reason, and thus GRBs, or at least a part, should be originated
within the effective region of the solar wind. 
GRBs may be possible from exploding electric double layers  
in instable plasma within the heliosphere (Li 1996).  
With respect to their isotropic spatial distribution, GRBs should be produced
far away from the Sun. 
A distance of about 100 AU can satisfy the constraint from observed GRB 
isotropy distribution (the measured dipole moment $\cos\theta\sim0.01$)
and  that from triangulation and direct positioning for a few bursts. 
At the solar-wind terminal shock,
near the heliospheric boundary and surrounding the solar system, lies an abrupt 
discontinuity, turbulent and complex structure (Suess 1990), 
at which various intriguing phenomena may occur, GRB might be one of them. 
The typical energy released in a GRB produced at a distance of 100 AU
is about $10^{26}f$ ergs, where $f$ is the beaming solid angle fraction.
If emission is beamed into an angular region of $\sim$ 1 deg$^{2}$, the 
typical energy of GRBs will be only $\sim 2\times10^{21}$ ergs,
equivalent just an average thunderstorm (Battan 1961). 

 A burst with total released energy $2\times 10^{21}$ erg  emits $\sim 2\times 
10^{28}$ photons (60 keV$/$ph assumed). If each discharge electron produces 
one burst photon through the inverse Compton scattering with thermal photons, a 
discharging current of $\sim 2\times 10^{8}$ A can produce the burst
emission with a 10 s duration.
A current flowing in the heliospheric plasma may contract by the magnetic
confinement and form a plasma cable with much larger density than the 
surroundings (Alfv\'{e}n 1981). In the disruptive discharge the cable is 
further pinched into a very narrow column.
The interaction length of the Compton scattering between energetic electrons
and thermal photons in a plasma column with diameter $d=1$ m and temperature 
$10^{6}$ K, which is pinched from 
a initial current cable with diameter $d_{0}=0.1$ AU$=1.5\times 10^{12}$ cm 
and density $N_{0}=1$ cm$^{-3}$, is only $\sim 10$ km. 
The density of the pinch column can be estimated as $N\sim 2\times 10^{20}$
cm$^{-3}$ and optical depth $\tau_{0}\sim 0.1$.
    
  The GRB spectra accumulated by the Spectroscopy Detectors (SD) of
BATSE are described well by the following empirical Band \etal (1993) model 
\begin{eqnarray}
N(E) &=& A\epsilon^{\alpha}\exp(-E/E_{0}) \hspace{31mm} 
(\alpha -\beta )E_{0}\geq E \nonumber \\
     &=& A[(\alpha -\beta )E_{0}]^{\alpha -\beta}\exp(\beta -\alpha )E^{\beta} 
\hspace{6mm} (\alpha -\beta )E_{0}\leq E \nonumber
\end{eqnarray}
\noindent with $\alpha\sim -1, \beta\sim -2$. The peak power energy 
$E_{pk}=(2+\alpha)E_{0}$. Our calculations show that $\gamma$-ray
flashes produced by large voltage and high temperature discharge 
plasma columns can also be described well by this model (Li \& Wu 1997).
We assumed a simple discharge model that an initial
discharge voltage $V_{0}$ is produced at time $t=0$, the discharging current
firstly has a Gaussian rise, $i(t)\propto \exp(-(\frac{t_{m}-t}{a_{1}})^{2})$
till to its maximum at $t=t_{m}$ and then falls exponentially with a decay
constant $a, i(t)\propto \exp(-\frac{t-t_{m}}{a})$; the potential difference
falls proportionally to the total quantity of charge transferred by the current,
$V(t)=V_{0}-B\int_{0}^{t} i(t)dt$. At time $t$, the kinematic energy of a 
discharge electron increases lineally along the path from 0 at the cathode 
to $V(t)$ at the anode,
the observed radiation is a summation of the scattered photons 
from electrons with kinematic energies between 0 and $V(t)$.
For example, let $V_{0}=100$ MV, $a_{1}=0.4a$, total discharge time $D=2.5a$,
the voltage at the end of discharge $V_{D}=5$ MV, the corresponding time
histories of discharge voltage and current are shown in Fig.12(a), where
the time is scaled by the current decay constant $a$. 
Differential scattered photon spectra from electrons with different energies
and isotropic thermal emission with $T=10^{6}$ K and $\tau_{0}=0.1$
were calculated by the Klein-Nishina formula.
Because the plasma instability, the discharge column may be twisting,
flickering and branching,
an observed spectrum should be an average over a certain interval of scattering
angle. As an example, the crosses in Fig.12(b) show the expected spectrum
observed by SD of BATSE for scattered photons with scattering angle 
$\leq 0.5$\degg, the error bars were estimated by assuming the total number
of recorded photons of $E\geq 20$ keV being $2\times 10^{4}$, a $E^{-2}$ 
background spectrum
with a flux of $1\times 10^{-2}$ cm$^{-2}$s$^{-1}$keV$^{-1}$ at 100 keV
and in considering the SD full-energy peak efficiency (Fishman \etal 1989).
Fitting the Band \etal model to the simulated spectrum in the region of $E\geq 20$ keV
we got the solid line in Fig.12(b) with $\alpha=-1.1$,
$ \beta=-3.2$, $E_{pk}=748.0$ keV and the reduced $\chi^{2}=0.59$,
showing the discharge model can produce a typical GRB spectrum.
     
\begin{figure}  %Fig.12
\psfig{figure=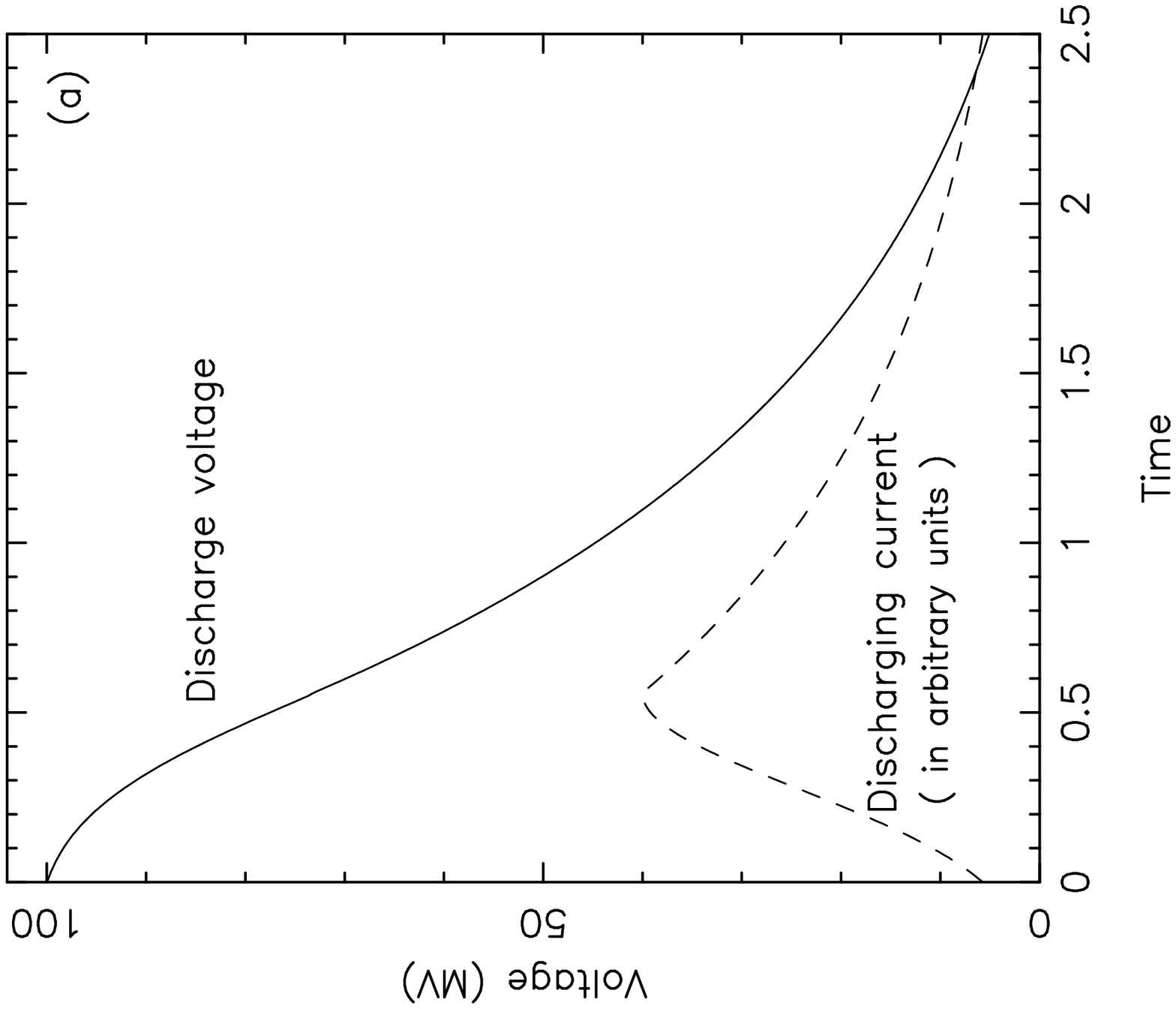,%
bbllx=88bp,bblly=18bp,bburx=600bp,bbury=500bp,%
height=7.0truecm,width=7.0truecm,angle=270}
\psfig{figure=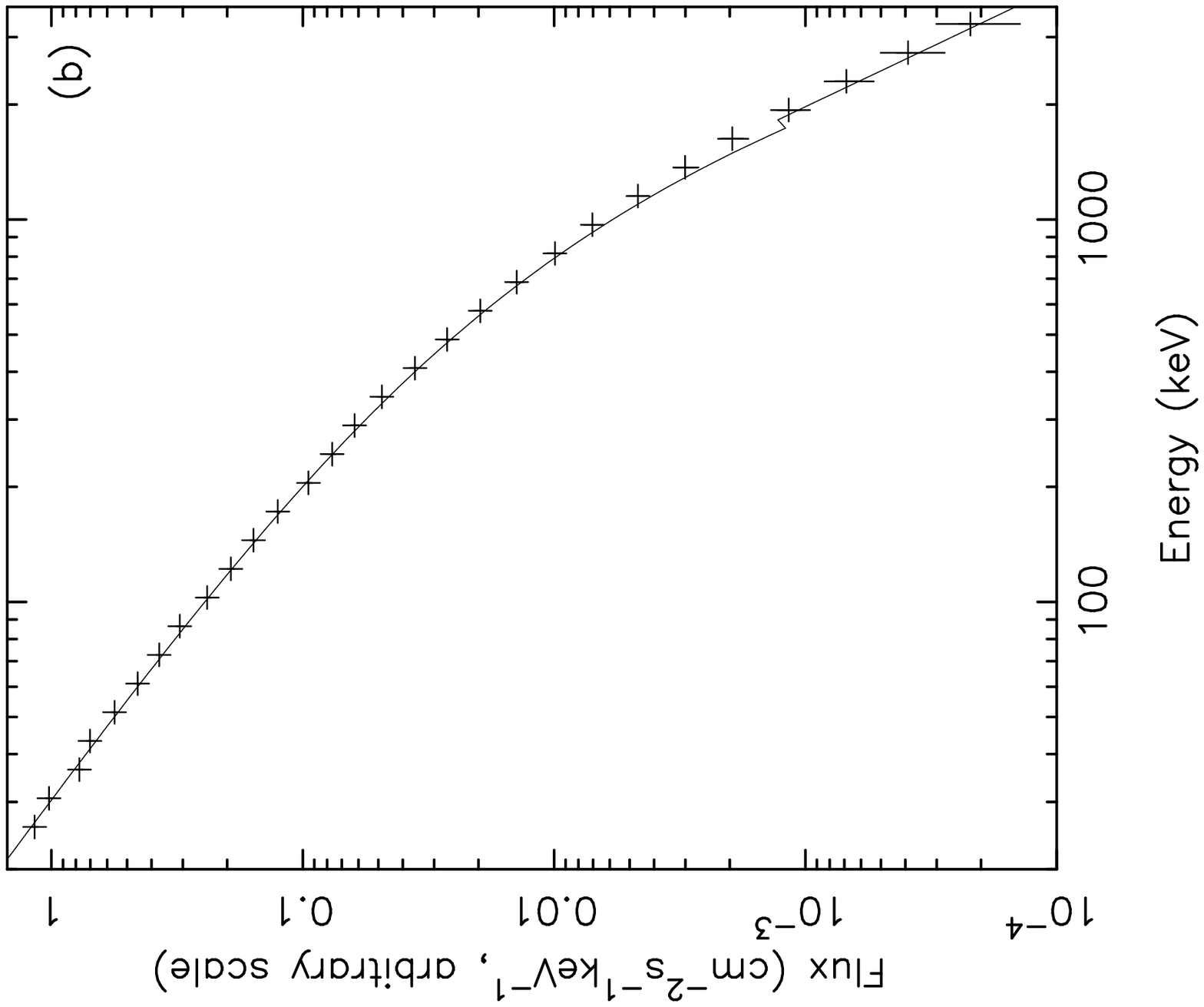,%
bbllx=88bp,bblly=28bp,bburx=600bp,bbury=500bp%
height=7.0truecm,width=7.0truecm,angle=270}
\caption[]{An example of discharge model. (a) Time histories of discharge
voltage (solid line) and current (dashed line), the voltage is in units of
$10^{6}$V, current in arbitrary units, time is scaled by the exponential 
decay constant of the current. (b) The calculated time-averaged spectrum 
(crosses) from a discharge column with temperature $T=1\times 10^{6}$ K,
optical depth at $1kT$ $\tau_{0}=0.1$ and
time histories of voltage and current shown in (a), observed 
over a scattering angle interval 0\degg -- 0.5\degg. 
The flux is in arbitrary scale. The solid line shows the Band \etal spectrum
with parameters $\alpha=-1.12$, $\beta=-2.28$ and $E_{pk}=826.4$ keV.  
   }
\end{figure}

  Producing Band \etal spectra is a general property of the discharge mechanism.
We calculated 100 spectra with different values of the discharge model
parameters. In our calculations, the initial voltage $V_{0}$ was sampled 
uniformly from the region
of $50-10000$ MV; the temperature of thermal emission, $T$, sampled from
$5\times 10^{5} - 5\times 10^{8}$ K and optical depth at $1kT$, $\tau_{0}$, 
from $0.01-1.0$ uniformly on a logarithmic scale;
the current decay constant was taken
as unit, $a=1$; the current rising constant $a_{1}$ was sampled from 
$0.01-0.7$; the voltage at the end of a discharge sampled from $0.5V_{0}-5$ MV;
the center of scattering angle region considered was taken randomly
from 0\degg -- 0.5\degg  and the width of the interval from 
0.1\degg -- 0.5\degg. For each
parameter set, we calculated the expected spectrum from the 
discharge process and fitted the Band \etal model  to. Most spectra 
fit well with the model, the average reduced $\chi^{2}=1.2\pm 0.3$.
Fig.13 is a plot of $\alpha$ vs. $E_{pk}$ from our calculations for 100 samples
of discharge model, which shows a distribution quite similar with that
from observed GRBs (see Preece \etal 1996, Fig.6). More calculations show that 
a summation
of several spectra from different discharge columns with different parameters
can also be well described by the Band \etal model.
 
\begin{figure}   %Fig.13
\psfig{figure=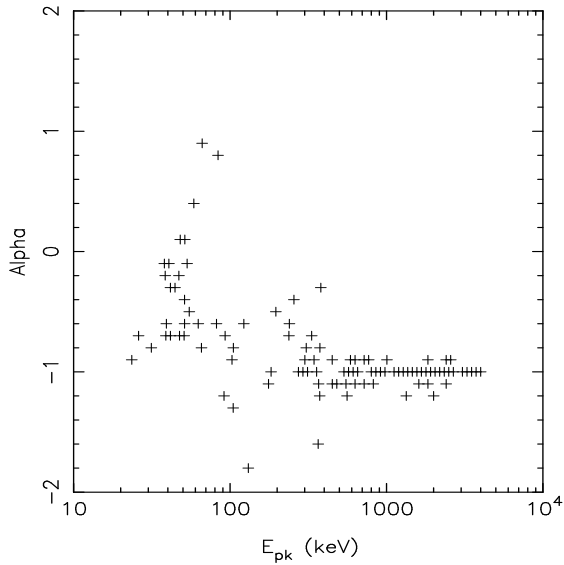,%
bbllx=40bp,bblly=28bp,bburx=600bp,bbury=500bp,%
height=8.0truecm,width=8.0truecm,angle=270}
\caption[]{Distribution of $\alpha$ and $E_{pk}$ from the fit of 
Band \etal model to calculated spectra for 100 samples of discharge
model.}
\end{figure}

   A hard-to-soft spectral evolution has been observed for many bright 
GRBs (Ford \etal 1995; Briggs 1995): the peak energy, $E_{pk}$, softens 
over the burst. Liang \& Kargatis (1996) further found that for a sample 
of bursts consisting
of well resolved, isolated pulses $E_{pk}$ decreases exponentially with 
photon fluence (running time integral of the flux), and that for many
multi-pulse bursts, the exponential decay constant is invariant from
pulse to pulse. The above properties can be interpreted easily in the frame
of discharge mechanism. For example, we calculated time-resolved spectra
for the discharge model described by Fig.12. To focus on the evolution
of $E_{pk}$, we followed Liang \& Kargatis (1996) to fix the two slopes 
throughout the
burst by fitting the Band \etal model to the time-average spectrum Fig.12(b),
the obtained time history of $E_{pk}$, Fig.14(a), shows a typical hard-to-soft 
evolution. In Fig.14(b) $E_{pk}$ is plotted against photon fluence
(the relative errors of $E_{pk}$ are simply taken as $10\%$, a typical value
for bright GRBs of BATSE), showing an exponential decay.
It is found from calculations that the decay constant of $E_{pk}$ 
versus fluence  is mainly determined by the velocity of discharge voltage
decaying, i.e. by the parameter $B$ of discharge circuit. Therefore,
the similarity of the exponential decay constants of the subsequent pulses
to that of the leading pulse in samples of multi-pulse bursts can be
explained as such pulses are from repeat discharges of the same circuit 
for the leading pulse. The softening trends in $E_{pk}$ is not a universal
property for the discharge model.
For the case of both the initial discharge voltage
and temperature being high, the cross section of Compton scattering falling
with the increase of electron energy (Akhiezer \& Berestetskii 1965)
causes $E_{pk}$ has not considerable
change or even increase with time. Ford \etal (1995) have already observed such 
evolution trend in a few bursts.
More dedicated model of discharge may explain
the phenomenon of envelope decaying observed in long multi-pulse
bursts (Ford \etal 1995).

\begin{figure} %Fig.14
\psfig{figure=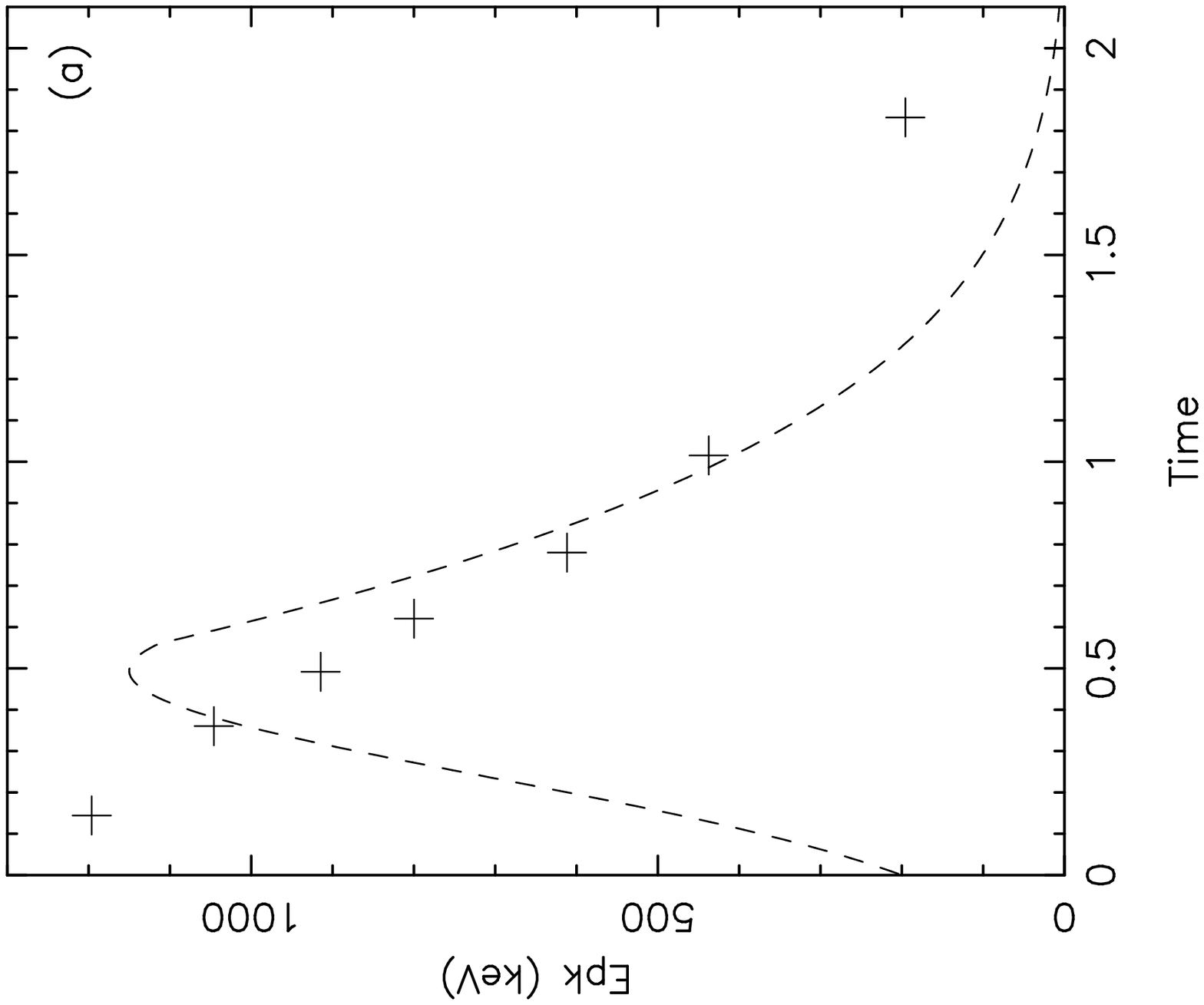,%
bbllx=88bp,bblly=18bp,bburx=600bp,bbury=500bp,%
height=7.0truecm,width=7.0truecm,angle=270}
\psfig{figure=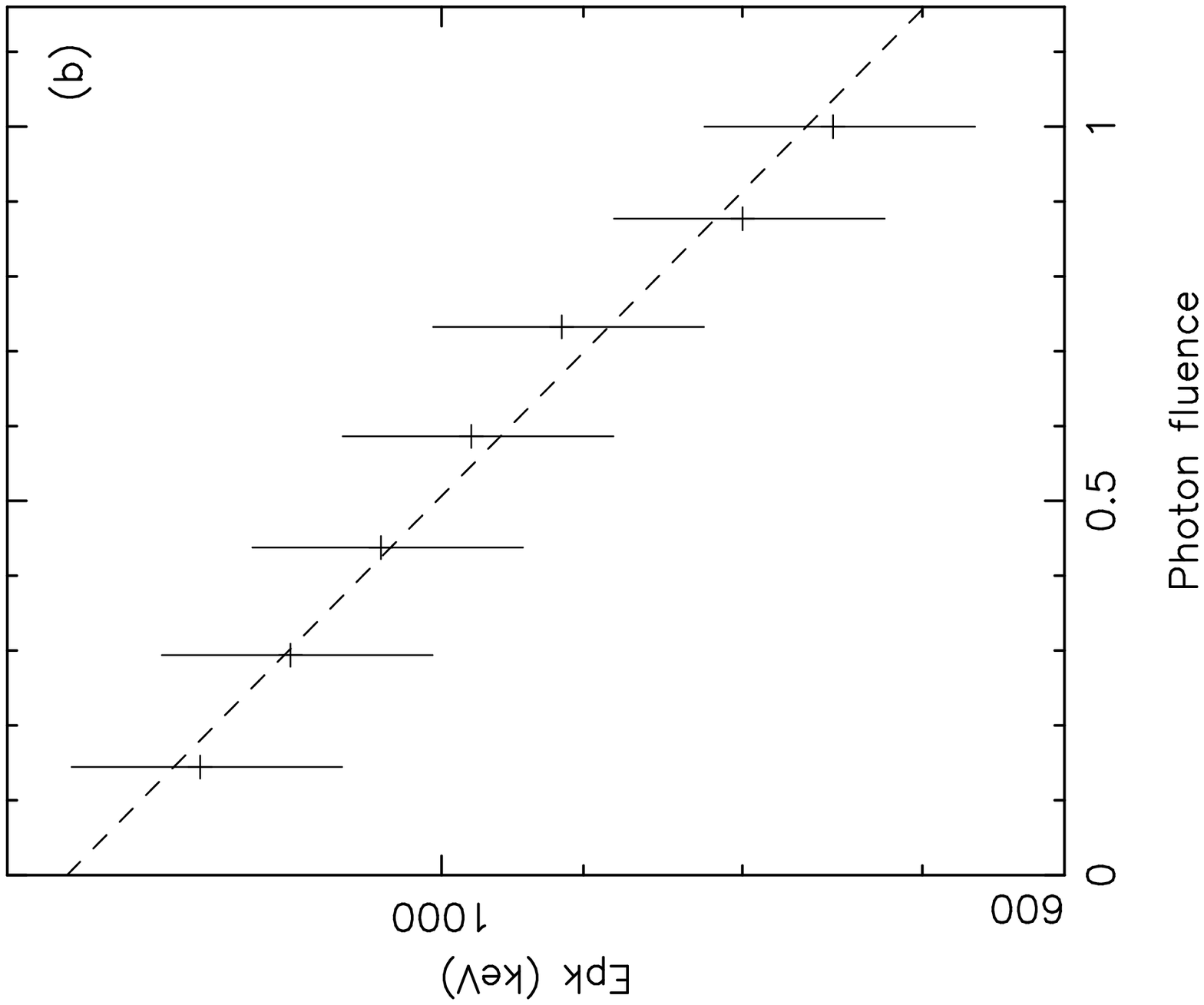,%
bbllx=88bp,bblly=28bp,bburx=600bp,bbury=500bp,%
height=7.0truecm,width=7.0truecm,angle=270}
\caption[]{Spectral evolution expected for the discharge model described
by Fig.12. (a) Time histories of $E_{pk}$ (keV) (crosses) and photon flux
(dashed line, in arbitrary units). The time is scaled by the decay time
constant of discharging current. (b) $E_{pk}$ versus photon fluence
in units of total fluence. Relative errors of $E_{pk}$ are taken as $10\%$.
Dashed line shows an exponential decay.   }
\end{figure}

  Another natural deduction of the discharge model is the durations becoming
shorter with increasing energy. For the discharge column described by Fig.12
we calculated the light curves in four energy channels: $20-50, 50-100, 
100-300$ and $>300$ keV, and their autocorrelation functions respectively.
The calculated autocorrelation function decreases more rapidly with time lag
in the higher energy channels compared to the lowest one,
as shown in Fig.15, which is consistent with that observed for GRBs (Link
\etal 1993).
We also calculated the cross-correlation functions between channel 1
($20-50$ keV) and 3 ($100-300$ keV) and found a maximum at the time lag
of $-0.01$ (in units of the time constant $a$ of current decay),
indicating a time delay between soft and hard emissions, which is
also observed in GRBs (Chipman 1993; Cheng \etal 1995).

\begin{figure}  %Fig.15
\psfig{figure=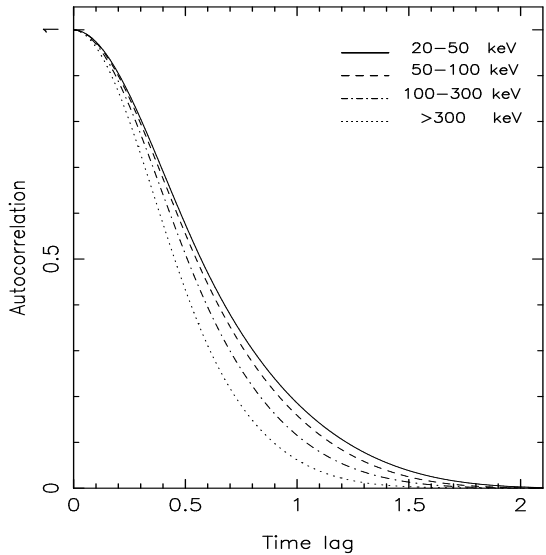,%
bbllx=40bp,bblly=28bp,bburx=600bp,bbury=500bp,%
height=8.0truecm,width=8.0truecm,angle=270}
\caption[]{Autocorrelation functions of light curves 
in four energy channels
expected by the discharge model of Fig.12 and Fig.14. The time lag
is scaled by the time constant of discharging current decay.}
\end{figure}

\subsection*{4. Discussion}

  The morphological diversity, the most striking feature of GRBs, is not
a difficulty for the discharge mechanism to explain. Complicated patterns, such as
wide variety of profile configurations, rich in fluctuation or smooth 
structures,
rapid rise vs. slower fall, weak precursor and secondary pulses etc, 
are common in various electrical disruptive 
discharges, e.g. various kinds of gas discharge, high vacuum disruption,
or lightning flashes. 
McBreen \etal (1994) have noted that there is considerable similarity in
the statistical properties of the light curves and peak flux distributions 
of GRBs and terrestrial lightning:
The range in durations of the flashes and multiple stroke patterns within 
the flashes are well fitted by lognormal distributions
and the integral number versus peak flux distributions by truncated lognormal
distributions.
Fishman \etal (1994a) discovered a dozen of brief, intense gamma-ray flashes
of atmospheric origin, terrestrial gamma-ray flashes (TGFs), from the first 
two years observation of BATSE.
Although TGFs have durations, typically a few 
milliseconds, much shorter than GRBs, their time profiles showing complex 
features in wide variety are quite similar with GRBs. TGFs have hard spectra
with the hardness ratio $H_{32}=F_{3}/F_{2}$ being about two times $<H_{32}>$
for GRBs. For all bursts, of which the energy fluences $F_{2}$ and $F_{3}$ are 
available in the 3B catalog, we found $<H_{32}>=3.82$. There exists an 
anti-correlation between GRB hardness and duration, hardness with short events 
in general being harder. Fig.16 shows the 
relation between the hardness ratio $H_{32}$ and duration $T_{90}$ in
the 3B bursts and the regression line 
$H_{32}=5.1-1.1\lg(T_{90}(s))$. Extrapolating above relation to 
$T_{90}=5\times10^{-3}$ s,
we expect $H_{32}=7.6$, just about two times $<H_{32}>$. 
Thus, besides the similarity in morphology, TGFs can 
be seen, in view of the spectral behaviour, as an extrapolation of GRBs.
An evident correlation of TGFs with storm system indicates that these events 
are caused by electrical discharges to the stratosphere or ionosphere of the 
Earth (Fishman \etal 1994a), then TGFs give an observational evidence 
that the discharge process can produce high-energy bursts like GRBs. 

\begin{figure} %Fig.16
\hbox{
\centerline{
\epsfig{figure=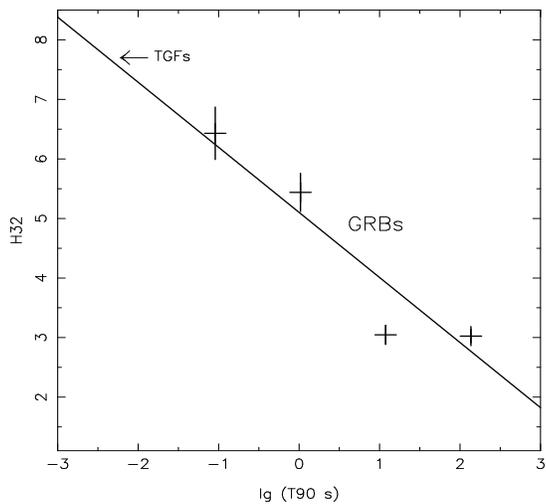,width=9.cm,height=10cm,angle=270,%
bbllx=200,bblly=18,bburx=774,bbury=594}
}}
\caption{Correlation between hardness and duration of GRBs in 3B catalog.
The arrow indicates the typical value of $H_{32}$ for terrestrial 
$\gamma$-ray flashes detected by BATSE.}
\end{figure}
\begin{figure}
\hbox{
\centerline{
\epsfig{figure=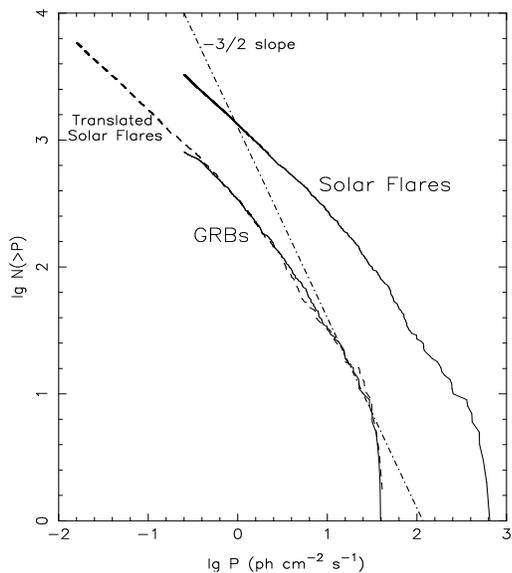,width=9.0cm,height=10.cm,angle=270, %
bbllx=200,bblly=18,bburx=774,bbury=594}
}}
\caption{Integral intensity distributions of GRBs and solar flares.
Peak fluxes $P$ of GRBs are measured on the 1.024s timescale
for 3B bursts, that of solar flares are taken from  
the BATSE solar flare catalog in SDAC for the period from 1991 April 19
to 1995 November 28. The dashed line is drawn 
from the solar flare $\lg N-\lg P$ distribution by a parallel translation    
$\lg N +0.25, \lg P-1.2$.}
\end{figure}

  Another kind of high-energy transients similar with GRBs is the solar hard
X-ray flare. It is not easy to make the distinction between GRBs and solar 
hard X-ray flares just by their time profiles. In burst recognition, 
solar flares 
distinguished from GRBs by their locations near the sun, their generally soft 
spectra, and their coincidence with solar X-ray events detected by the GOES
satellite (Fishman \etal 1994b). Opposite TGFs, solar flares are in general
longer and softer than GRBs. 
From 1991 April 19 to 1995 November 28, there were a total of 4467 solar 
flares detected by BATSE, of which 677 gave rise to a burst trigger (SDAC 1996).
Yu \etal (1997) have found that the intensity distribution of BATSE solar
flares is very similar with that of GRBs. Fig.17 shows the intensity 
distribution
of solar flares in the BATSE flare catalog  and that of GRBs in the 3B catalog.
The solar flare integral $\lg N-\lg P$ distribution also appears a
$-3/2$ slope for bright events and saturated at low intensities like GRB
does. After parallel translation the $\lg N-\lg P$ diagram of solar flares
is well coincidence with that of GRBs. This remarkable fact implies that the
characteristic of GRB intensity distribution is not necessarily to be a
measure of spatial homogeneity of burst sources, and that GRBs and solar 
flares may have similar production mechanism. 
Another noticed similarity between GRBs and solar flares is 
the delayed 1-GeV emission observed in both events (Kanbach \etal 1993;
Hurley \etal 1994).     
In fact, Alfv\'{e}n and Carlqvist (1967) suggested exploding discharges of
electric double layers to be responsible for solar flares. 

\parindent=0mm
{\sl Acknowledgements}. This research has made of data obtained through
the Compton Observatory Science Support Center and the Solar
Data Analysis Center at the
NASA-Goddard Space Flight Center. The author thanks Professors Lu Tan and 
Wu Mei for helpful discussions. This work was supported by the National Natural 
Science Foundation of China.  

\vspace{4mm}
%\newpage
{\bf References}
\vspace{2mm}
\parindent=0mm
\parskip=0mm
\begin{verse}

Akhiezer A.I. \& Berestetskii V.B., 1965, Quantum Electrodynamics,
John Wiley \& Sons, Inc. 

Alfv\'{e}n H., 1981,  Cosmic Plasma, D.Reidel Publishing Company

Alfv\'{e}n H. and Carlqvist P., 1967, Solar Phys. 1, 220 

Band D.L. \ital 1993, ApJ 413, 281

Battan L.J., 1961, The Nature of Violent Storms, Doubleday, New York, 21

Briggs M.S., 1995, in Ann. New York Acad. Sci. 

Cheng L.X., Ma Y.Q., Cheng K.S., Lu T. \& Zhou Y.Y., 1995, A\&A 300, 746

Chipman E., 1993, in AIP Conf. Proc. 307: Gamma Ray bursts, Huntsville,
p.202

Dermer C.D., 1992, Phys. Rev. Letters 68, 1799

Dezalay J.P. \ital 1992, in AIP Conf. Proc.265, Gamma-Ray Bursts, 

Diaconis P. \& Efron B., 1983, Scien. American, Sept. p.96

Efron B. 1979, Ann. Statistics 7, 1

Fishman G. J. \ital 1989, in GRO science Workshop, Greenbelt, 2-39

Fishman G.F. \ital 1994a, Science 264, 1313

Fishman G.F. \ital 1994b, ApJS 92, 229

Ford L.A. \ital 1995, ApJ 439, 307

Ginzburg V.L. \& Syrovatskii S.I., 1964,  The Origin of Cosmic Rays, 
Pergamon Press

Hartmannm D.H., 1994, in G. Fishman, J. Brainerd \& K. Hurley, eds.
Gamma-Ray Bursts, 2nd Workshop, AIP Press, New York

Hurley K. \ital 1994, Nature 372, 652

Kanbach G. \ital 1993, A\&AS 97, 349

Klebesadel R., Strong I.B. and Olson R.A., 1973, ApJ 182, L85

Kouveliotou C. \ital 1993a, ApJ 413, L101

Kouveliotou C. \ital 1993b, A\&AS 97, 55

Krall N.E. \& Trivelpiece A.W., 1973,  Principles of Plasma Physics,
McGraw-Hill, Inc.

Kulsrud R.M., 1954, ApJ 119, 386

Lamb D.Q., 1995, PASP 107, 1152

Liang E. \& Kargatis V., 1996, Nature 381, 49
 
Li T.P., 1996, Chin. Phys. Lett. 13, 637

Li T.P., 1997, Acta Astrophys. Sinica, in press

Li T.P. \& Wu M., 1997, Chin. Phys. Lett., in press

Link B., Epstein R.I. \& Priedhorsky W.C., 1993, ApJ 408, L81

Mao S. \& Paczy\'{n}ski B., 1992, ApJ 388, L45
 
McBreen B., Hurley K.J., Long R. \& Metcalfe L., 1994,  MNRAS 271, 662

Meegan C.A., Fishman G.J., Bhat N.D. \ital 1996, The Third BATSE 
Gamma-Ray Burst Catalog, ApJS 106, 65

Paczy\'{n}ski B., 1995, PASP 107, 1167

Piran T., 1992, ApJ 389, L45

Preece R. D. \ital 1996, ApJ 473, 310

SDAC, 1996, BATSE flare catalog, www address:\\
\hspace{5mm} http://umbra.nascom.nasa.gov/batse/batse\_years.html

Sieger S., 1956, Nonparametric Statistics for The Behavioral Sciences,
McGraw-Hill Book Comp., New York

Suess S.T., 1990, Rev. Geophys., 28, 97
 
Yu W.F., Li T.P., Wu M. \& Song L.M., 1997, Acta Astrophys. Sinica, 17, 66
\end{verse}
\end{document}